\documentclass[sigconf]{acmart}

\usepackage{acmart-taps}
\usepackage[most]{tcolorbox}
\usepackage{amsfonts}
\usepackage{microtype}
\usepackage{balance}

\usepackage{textcomp}
\usepackage{xcolor}

\usepackage{color, colortbl}
\usepackage{subcaption}
\usepackage[utf8]{inputenc}
\usepackage{verbatimbox}
\usepackage{multirow}
\usepackage{makecell}
\usepackage[scientific-notation=true]{siunitx}
\usepackage{todonotes}
\usepackage{balance}

\usepackage{ dsfont }
\usepackage[Algorithm,ruled]{algorithm}
\usepackage{url}
\usepackage[noend]{algpseudocode}

\usepackage{textcomp}
\usepackage{xcolor}
\usepackage{xspace}
\usepackage{enumitem}
\usepackage{listings}
\usepackage{pifont}
\usepackage{listings,multicol}
\usepackage{float}
\usepackage[utf8]{inputenc}
\usepackage{verbatim}
\usepackage{multirow}
\usepackage{makecell}
\usepackage[scientific-notation=true]{siunitx}
\usepackage{balance}
\usepackage{amsmath,amsfonts}

\usepackage[Algorithm,ruled]{algorithm}
\usepackage{url}
\usepackage[noend]{algpseudocode}
\usepackage{hyperref}

\usepackage{textcomp}
\usepackage{xcolor}
\usepackage{booktabs}
\usepackage{xspace}
\usepackage{enumitem}
\usepackage{listings}
\usepackage{pifont}
\usepackage{framed}
\usepackage{xcolor}

\definecolor{codered}{HTML}{D9534F}
\definecolor{codegreen}{HTML}{5CB85C}
\definecolor{codeblue}{HTML}{0000FF}

\definecolor{m1}{RGB}{224, 250, 255}
\definecolor{m2}{RGB}{219, 224, 255}
\definecolor{m3}{RGB}{252, 255, 220}
\definecolor{m4}{RGB}{249, 214, 177}

\usepackage{amsthm}
\newtheorem{definition}{Definition}
  
\usepackage{tikz}
\usetikzlibrary{shapes.geometric}
\usepackage{listings}
\usepackage{pgfplots}
\usepackage[]{footmisc}

\tikzset{
    use bounding box relative coordinates/.style={
        shift={(current bounding box.south west)},
        x={(current bounding box.south east)},
        y={(current bounding box.north west)}
    },
}

\tikzstyle{normal} = [rectangle,text centered,draw=white]

\copyrightyear{2026}
\acmYear{2026}
\setcopyright{cc}
\setcctype{by}
\acmConference[FORGE '26]{2026 IEEE/ACM Third International Conference on AI Foundation Models and Software Engineering}{April 12--13, 2026}{Rio de Janeiro, Brazil}
\acmBooktitle{2026 IEEE/ACM Third International Conference on AI Foundation Models and Software Engineering (FORGE '26), April 12--13, 2026, Rio de Janeiro, Brazil}
\acmPrice{}
\acmDOI{10.1145/3793655.3793727}
\acmISBN{979-8-4007-2477-0/2026/04}

\begin{document}

\title{Assessing, Exploiting, and Mitigating Syntactic Robustness Failures in LLM-Based Code Generation }

\author{Laboni Sarker}
\affiliation{%
  \department{Computer Science}
  \institution{University of California, Santa Barbara}
  \city{Santa Barbara}
  \state{CA}
  \country{USA}
}
\email{labonisarker@ucsb.edu}

\author{Mara Downing}
\affiliation{%
    \department{Computer Science}\institution{University of California, Santa Barbara}
  \city{Santa Barbara}
  \state{CA}
  \country{USA}
}
\email{maradowning@ucsb.edu}

\author{Achintya Desai}
\affiliation{%
    \department{Computer Science}\institution{University of California, Santa Barbara}
  \city{Santa Barbara}
  \state{CA}
  \country{USA}
}
\email{achintya@ucsb.edu}

\author{Tevfik Bultan}
\affiliation{%
    \department{Computer Science}\institution{University of California, Santa Barbara}
  \city{Santa Barbara}
  \state{CA}
  \country{USA}
}
\email{bultan@ucsb.edu}

\begin{abstract}
Rapid advances in the field of Large Language Models (LLMs) have made LLM-based code generation an important area for investigation. An LLM-based code generator takes a prompt as input and produces code that implements the requirements specified in the prompt. Many software requirements include mathematical formulas that specify the expected behavior of the code to be generated.
Given a code generation prompt that contains a mathematical formula, a reasonable expectation is that, if the formula is syntactically modified without changing its semantics, the generated code for the modified prompt should be semantically equivalent. 
We formalize this concept as syntactic robustness and investigate the syntactic robustness of LLMs as code generators.
Our experimental assessment demonstrates that LLMs are not syntactically robust for code generation prompts with formulas, especially for the ones that require mathematical reasoning.
We investigate attack strategies that can further
deteriorate the syntactic robustness of LLMs.
Finally, to mitigate syntactic robustness failures in LLMs, we propose a pre-processing step that uses reductions to transform formulas in prompts to a simplified form. Our experimental results demonstrate that the syntactic robustness of LLM-based code generation improves significantly using our approach,
improving syntactic robustness of LLMs from 54.05\% to 74.42\%.

\end{abstract}



\maketitle


\section{Introduction}
\label{sec:introduction}

\let\thefootnote\relax
\footnotetext{This material is based on research supported by NSF under grants 2124039 and 2008660 and DARPA under the agreement number N66001-22-2-4037. The U.S. Government is authorized to reproduce and distribute reprints for Governmental purposes notwithstanding any copyright notation thereon. The views and conclusions contained herein are those of the authors and should not be interpreted as necessarily representing the official policies or endorsements, either expressed or implied, of the U.S. Government.}

Large language models (LLMs) are becoming popular in the software engineering community~\cite{fan2023large,LLM4SE}, especially for code generation tasks~\cite{surveyCodeGeneration,jiang2023self,robustnessCopilot,recode}, and have shown strong performance on code generation benchmarks~\cite{huang2023agentcoder,sotapaperswithcode,LLMTrainedonCode}. 
Recently, LLMs have also been used for code generation in the scientific computing domain~\cite{kashefi2023chatgpt,llm4ScientificDiscovery,shojaee2024llm,du2024llm4ed,ChatGptForDE, ChatGptForDEpresentation, GptForFiniteElementMethod} and
for generating code that efficiently handles complex mathematical computations~\cite{ChatGptForDE, ChatGptForDEpresentation, GptForFiniteElementMethod,vishnu2025unveiling,coignion2024performance}. Furthermore, LLMs have shown promise in handling formal specifications ~\cite{formalProgramSynthesisLLM, clover}, which rely on mathematical formulas in the specification of software engineering tasks. As its adoption grows, there is a critical need for systematic evaluation of the correctness and reliability of LLM-based code generation~\cite{fan2023large,robustnessCopilot,recode,jiang2022discovering,yan2023coco,stackOverReplacementRobustness,buscemi2023comparative,xu2022systematic}.
Reliability in machine learning systems is often assessed through robustness, which has been extensively studied for classification tasks~\cite{katz2019marabou,bunel2020branch,sun2018concolic,singh2019abstract,wang2018formal,xie2019deephunter,baluta2021scalable}. Robustness in classification models involves defining a neighborhood around an input with a known classification and ensuring the same classification is obtained within that neighborhood~\cite{gehr2018ai2}. For generative models, however, this definition must be adapted, as obtaining and verifying objectively accurate outputs is more complex---many different outputs may be objectively correct. This highlights the need for well-defined robustness metrics to analyze LLM-based code generators, which are a subdomain of generative models.

Several prior works~\cite{robustnessCopilot,jiang2022discovering,yan2023coco,stackOverReplacementRobustness,buscemi2023comparative,xu2022systematic,recode,fan2023large} highlight the importance of studying robustness for code generation by LLMs, and some do investigate robustness~\cite{yan2023coco,recode,nondeterminism}. However, the impact of syntactic transformations of the prompts together with the increasing syntactic distance (Section~\ref{sec:transformation_1}) has not been investigated. Additionally, black-box attacks specifically designed to expose failures in the robustness of LLM-based code generators in handling syntactic variations of semantically equivalent prompts (Section~\ref{sec:attack}) have not been explored.

In this paper, we define a formal robustness measure for LLM-based code generators called syntactic robustness (Section~\ref{sec:robustness}) and analyze the performance of LLMs with respect to this measure (Section~\ref{sec:experiments}). Intuitively, we define syntactic robustness as the degree to which the semantically equivalent prompts elicit semantically equivalent responses from the LLM code generator.

We define a set of syntactic transformations (Section~\ref{sec:transformation_1}) that generate syntactically different but semantically equivalent variations of mathematical formulas to assess the syntactic robustness of LLM-based code generators. We present attack strategies (Section~\ref{sec:attack}) to exploit the syntactic robustness failures of the existing LLMs while minimizing the modifications to the formulas in the prompts. 
Furthermore, we present a pre-processing step (Section~\ref{sec:prepocessing}) to mitigate syntactic robustness failures and defend against such attacks, and demonstrate its effectiveness. Our assessment reveals that state-of-the-art LLMs lack syntactic robustness and that prompts requiring mathematical reasoning exhibit lower robustness compared to those involving direct translation of mathematical formulas into code. Our experimental evaluation also shows that our attacks can effectively exploit the syntactic robustness failures of the LLMs.
Finally, our experiments demonstrate that our pre-processing effectively mitigates syntactic robustness failures, significantly improving syntactic robustness of LLMs.

\textit{Contributions.} Our contributions in this paper are as follows:
\begin{enumerate}
    \item A formal definition of syntactic robustness (Section~\ref{sec:robustness}).
     \item A set of prompts with mathematical formulas, designed to benchmark the robustness of LLM-based code generators (provided as artifact~\cite{ourAnonRepo}).
    \item A set of rules to mutate the mathematical formulas, that along with the original prompts, can be used for syntactic robustness evaluation (Section~\ref{sec:transformation_1}).  
    \item A set of black-box attacks for exploiting syntactic robustness failures of LLMs (Section~\ref{sec:attack}). 
    \item A prompt pre-processing technique that helps in mitigating syntactic robustness failures of LLMs (Section~\ref{sec:prepocessing}).  
    \item Empirical assessment of syntactic robustness, along with novel attack and mitigation techniques, on state-of-the-art LLM foundational models and frameworks (Section~\ref{sec:experiments}).
\end{enumerate}


\textit{Organization.} We discuss motivating examples of code generation prompts in Section~\ref{sec:overview}, then we provide a formal definition of syntactic robustness and related concepts in Section~\ref{sec:robustness}. We discuss our formula transformation procedure in Section~\ref{sec:transformation_1} and detail our attacks in Section~\ref{sec:attack}. We explain our pre-processing procedure in Section~\ref{sec:prepocessing}, and then explicate implementation design details in Section~\ref{sec:implementation}. We present our experimental evaluation in Section~\ref{sec:experiments}, discuss the related work in Section~\ref{sec:relwork}, and conclude the paper in Section~\ref{sec:conclusion}.

\section{Motivation}
\label{sec:overview}

Mathematical formulas and constraints are common in code generation prompts in a variety of domains~\cite{kashefi2023chatgpt, llm4ScientificDiscovery,vishnu2025unveiling,coignion2024performance,clover}. Given many equivalent ways one can write a mathematical formula or constraint, ensuring syntactic robustness is crucial for code generation in such contexts. Thus, it is critical to evaluate the robustness of LLM-based code generators for prompts with mathematical formulas. Note that, the equivalence of mathematical formulas can be defined clearly and unambiguously, allowing for an automated evaluation of  robustness~\cite{liu2021practical,hort2025semantic,tian2023code}. This is in contrast to prior works that mutate the text of the prompt~\cite{recode,robustnessCopilot}, which can alter the meaning due to the ambiguity of natural language.













\tcolorbox[colback=gray!10,colframe=black,width=\linewidth,arc=2mm, boxrule=.2mm]
\small {\bf Prompt A} {\em ``Implement a C program which takes `a',`b' as inputs where
method func(a: array<int>, b: array<int>)

returns (result: bool) 

requires a != null \&\& b != null 


ensures result ==>  exists i, j :: 0 <= i < a.Length \&\& 

0 <= j < b.Length \&\&  \textit{\textbf{\textcolor{blue}{Eq}}}

ensures !result ==>  forall i, j :: 0 <= i < a.Length \&\&

0 <= j < b.Length ==> a[i] != b[j].''}
\label{promptA_eq}
\endtcolorbox
\vspace*{-1.1em}
\tcolorbox[colback=gray!10,colframe=black,width=\linewidth,arc=2mm, boxrule=.2mm]
\small {\bf Expressions for placeholder {\textit{\textbf{\textcolor{blue}{Eq}}}} in Prompt A}

\small {\textit{\textbf{\textcolor{blue}{Eq}}}: a[i] == b[j]

\small \textit{\textbf{\textcolor{blue}{Eq}}}: (a[i]) * 7 - (b[j]) * 7 == (b[j]) * 7 - (b[j]) * 7}
\label{promptA_eq_equations}
\endtcolorbox
Syntactically transformed prompts with mutated formulas can also be interpreted as adversarial samples for assessing the robustness of LLM-based code generators under adversarial attacks~\cite{zhu2023promptrobust,adversarialAttackCodeSum,DLRobustness}. Our evaluation demonstrates that LLMs are vulnerable to adversarial prompt-based attacks, motivating the need for assessing and improving the syntactic robustness of LLM-based code generators. 
\begin{figure}[h]
    \centering
\includegraphics[width=.72\linewidth]{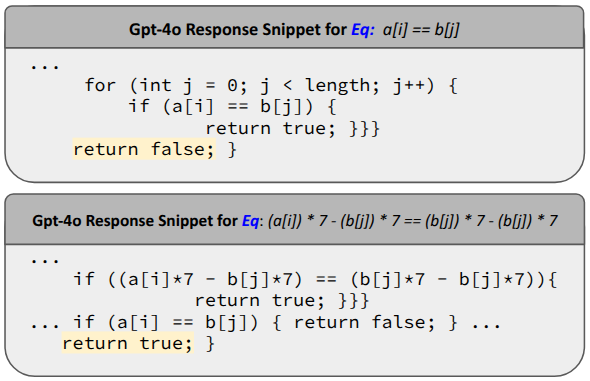}
    \caption{Code segments from the programs generated by Gpt-4o for Prompt A}
    \Description{Code segments of the responses generated by Gpt-4o for Prompt A}
    \label{fig:promptANewResult}
\end{figure}
\paragraph{Example prompts}
We motivate our work through a series of example code generation prompts involving mathematical formulas from different datasets. Prompt A is from the clover dataset~\cite{clover}, which contains formal specifications, Prompt B is generated from a HackerRank problem~\cite{HackerRank}, and Prompt C and D are from the MaTT dataset~\cite{MaTTBenchmarkPaper}. 

\begin{figure}[h]
    \centering
\includegraphics[width=.72\linewidth]{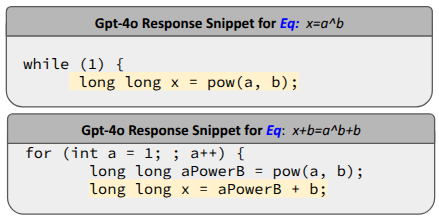}
    \caption{Code segments from the programs generated by Gpt-4o for Prompt B}
    \Description{Code segments of the responses generated by Gpt-4o for Prompt B}
    \label{fig:promptAResult}
\end{figure}

Prompt A and B demonstrate the syntactic robustness challenges in LLM-based code generation. Prompt A shows a code generation query to generate a program that returns \textit{True} if two input arrays share any common element and \textit{False} otherwise. On the other hand, Prompt B asks to generate a program that can find all b-digit numbers that are also the b-th power of a natural number. The placeholder \textit{\textbf{\textcolor{blue}{Eq}}} replacements for both Prompt A and B are shown at the bottom of each prompt and the corresponding outputs are presented in Figures~\ref{fig:promptANewResult} and~\ref{fig:promptAResult}, respectively.
\tcolorbox[colback=gray!10,colframe=black,width=\linewidth,arc=2mm, boxrule=.2mm]
\small {\bf Prompt B} {\em ``The number of digits present in a number is represented by `d'. `a' represents a natural number.  Write a C code which will return all possible `x' given a value `b' which will satisfy the following pre and post-condition. Pre-condition: 1 <= b  \&\& b <= 19 Post condition: d=b and \textit{\textbf{\textcolor{blue}{Eq}}}. Print only the value of all possible values of x in comma separated form.''}
\label{promptA_new}
\endtcolorbox
\vspace{-1.1em}
\tcolorbox[colback=gray!10,colframe=black,width=\linewidth,arc=2mm, boxrule=.2mm]
\small {\bf Expressions for placeholder {\textit{\textbf{\textcolor{blue}{Eq}}}} in Prompt B}

\small {\textit{\textbf{\textcolor{blue}{Eq}}}: x = a\textasciicircum b

\small \textit{\textbf{\textcolor{blue}{Eq}}}: x+b = a\textasciicircum b+b}
\label{promptA_new}
\endtcolorbox
Figures~\ref{fig:promptANewResult} and~\ref{fig:promptAResult} show code segments generated by GPT-4o highlighting the differences for semantically equivalent prompts. Despite the semantic equivalence of the mathematical formulas in the prompts, the generated codes reflect different interpretations.
Both examples demonstrate the case where Gpt-4o struggles to handle syntactic mutations on a simple equation. Note that, in both prompts, the natural language texts are kept the same---the only distinction is the semantically equivalent but syntactically different equations.
However, the generated codes from Gpt-4o are not semantically equivalent. In our experiments (Section~\ref{sec:experiments}), we show that this problem exists across multiple LLMs for a variety of prompts. 
\tcolorbox[colback=gray!10,colframe=black,width=\linewidth,arc=2mm, boxrule=.2mm]
\small {\bf Prompt C} {\em ``Newton's Law of Gravitation states that two bodies with masses a and b attract each other with a force F=9.8 * a * b / d\textasciicircum 2 where d is the distance between the bodies and 9.8 is the gravitational constant. If one of the bodies is fixed, implement a C program to find the work needed to move the other from d=c to d=e where `a', `b',`c',`e' are inputs to the program. Assume none of `a', `b', `c', `d' or `e' are 0. Print only the work needed up to 2 digit precision after decimal (do not print anything else).''}
\endtcolorbox

In code generation tasks, semantically equivalent formulas can often appear in syntactically different forms, particularly when derived from complex specifications. As the examples above demonstrate, enhancing the reliability of LLMs requires investigating and improving their syntactic robustness before deploying them. 

\paragraph{Reasoning and Translation Prompts}
We define two types of prompts for our analysis: reasoning and translation prompts. Consider the Prompt C below which asks for a program to find out the work done when a body is moved from one fixed position to another. The expected output from this prompt is to get the amount of work done which is $9.8 \times a \times b \times (1/c - 1/e)$ obtained by computing the integral $\int_c ^e (9.8 \times a \times b / d ^ 2) $ with respect to $d$. 


We create a variation of Prompt C (we call Prompt D) by replacing the second sentence of the prompt as
{\em 
``Implement a C program to find the attraction force where `a', `b' and `d' are inputs to the program.''}
Prompt D contains the same equation as Prompt C but asks to calculate the attraction force given the equation, where the expected output is computed by evaluating the  expression $9.8 \times a \times b / d^2$, which is provided in the prompt.


Although both prompts involve the same mathematical formula, Prompt C requires mathematical reasoning to perform integration, while Prompt D requires translation of the input formula to an expression in the target programming language to be evaluated during program execution. 
We name these two classes of prompts as {\em reasoning} and {\em translation} prompts, respectively. 
Our experiments indicate that syntactic robustness of LLM-based code generators differ for reasoning and translation prompts.

\section{Syntactic Robustness}
\label{sec:robustness}
In this section, we formalize the concept of syntactic robustness. We start by defining key concepts including LLM-based code generators, prompts, and generated programs and their equivalence, before introducing our formal definition of syntactic robustness.
\paragraph{LLM-based Code Generators}
We define an {\em LLM-based code generator} as follows:  
\begin{definition}
\small
An {\bf LLM-based code generator} $G$
takes a prompt $P \in \mathcal{P}$ as input and generates code $C \in \mathcal{C}$, denoted as
$G: \mathcal{P} \rightarrow \mathcal{C}$, where $\mathcal{P}$ is the set of prompts and $\mathcal{C}$ is the set of programs.
\end{definition}

In this paper, we focus on prompts that contain both English text and mathematical formulas:
\begin{definition}
\small
A {\bf prompt} $P \in \mathcal{P}$ consists of an English text $T$ and a mathematical formula $F$ denoted as a tuple $P \langle T, F \rangle$. 
\end{definition}
When an LLM-based code generator generates code $C$ for prompt $P\langle T, F\rangle$, we denote it as:
$G (P\langle T, F\rangle) = C$.

\paragraph{Semantic Equivalence of Formulas}
Two syntactically different formulas $F_1$ and $F_2$ can be semantically equivalent. We denote the semantic equivalence of two formulas $F_1$ and $F_2$ as: 
\[
 [\![ F_1 ]\!]  \equiv  [\![ F_2 ]\!] 
\]
which means that given the same valuation, formulas $F_1$ and $F_2$ evaluate to the same value. 
For example, consider the two formulas $F_1$ and $F_2$:
\[ F_1: a \times x + b = 0 \ \ \ \ \ \ \ \ \  
F_2: a\times x + a + b = a 
\]
Note that although $F_1$ and $F_2$ are syntactically different formulas
they are semantically equivalent
(i.e., $F_1 \neq F_2$ and
 $[\![ F_1 ]\!]  \equiv  [\![ F_2 ]\!]$).

\paragraph{Programs as Functions}
We focus on programs that can be modeled as functions. We assume that, for a given input, each program execution terminates and produces the same output. Formally:
\begin{definition}
\small
A  {\bf program} $C$ is a total function from the domain of inputs to the domain of outputs, $C: I \rightarrow O$, 
where 
$C(i) = o$ denotes that on input $i \in I$, the output of $C$ is $o \in O$.
\label{def:program}
\end{definition}

\paragraph{Program Equivalence}
We define equivalence of programs based on their input-output behavior:
\begin{definition}
\small
Given two programs $C_1: I_1 \rightarrow O_1$ and $C_2: I_2 \rightarrow O_2$ where  $I_1 = I_2$,
\begin{itemize}
\item 
$C_1$ and $C_2$ are {\bf equivalent}, denoted as
$[\![ C_1 ]\!] \equiv
[\![ C_2 ]\!]$,
if and only if, \ 
$\forall i \in I, C_1(i)=
C_2(i)$.
\item $C_1$ and $C_2$ are {\bf non-equivalent}, denoted as
$[\![ C_1 ]\!] \not \equiv
[\![ C_2 ]\!]$,
if and only if, \  
$\exists i \in I, C_1(i) \neq
C_2(i)$.
\end{itemize}
\label{def:equivalence}
\end{definition}
\noindent
Note that different implementations of the same functionality are considered equivalent according to this definition as long as the input-output behavior is the same. 

\paragraph{Syntactic Robustness}
We can now begin to define syntactic robustness for LLM-based code generators:
\begin{definition}
\small 
An LLM-based code generator $G$ is {\bf syntactically robust}, if and only if, 
given any two prompts $P\langle T,F_1\rangle, P\langle T,F_2\rangle \in \mathcal{P}$ where
$ [\![ F_1 ]\!]  \equiv  [\![ F_2 ]\!] $,
$[\![  G(P\langle T,F_1\rangle) ]\!]
\equiv [\![  G(P\langle T,F_2\rangle) ]\!]$.
\label{robustness}
\end{definition}
\noindent
i.e., an LLM-based code generator is syntactically robust if it generates equivalent code for semantically equivalent but  syntactically different prompts.  

There are two issues with the above definition that we will address. First, according to the definition, if an LLM-based code generator generates the same code for all prompts, it would be syntactically robust. i.e., an LLM-based code generator that for all prompts generates the same trivial code:

{\small 
\begin{verbatim}
int main() { printf("0\n"); return 0; }
\end{verbatim}
}

\noindent
would be syntactically robust. 
To address this problem, we introduce the concept of a reference code for each prompt as follows:
\begin{definition}
\small
Given a prompt $P \langle T, F \rangle$ we call
$R(P\langle T, F\rangle) = C^R_F$ the {\bf reference code} for the 
prompt $P \langle T, F \rangle$, where $C^R_F$ is a correct implementation of the requirements specified in the prompt $P \langle T, F \rangle$. 
\end{definition}
\noindent
Note that $C^R_F$ can be written manually or can be generated by a code generator and validated by different means (such as manual inspection, testing, or verification).

Second, Definition~\ref{robustness} of syntactic robustness requires the LLM-based code generator to generate semantically equivalent programs for all semantically equivalent prompts. Even if the code generator generates semantically different code for only one syntactically different prompt while generating semantically equivalent code for all other prompts, it is not syntactically robust according to the above definition. So, we extend our definition below to measure the {\em syntactic robustness degree}, where Definition~\ref{robustness}
corresponds to the highest degree of syntactic robustness. Syntactic robustness degree for an LLM-based code generator
for a given prompt, its reference code, and its syntactic variations as follows:
\begin{definition}
\small 
Given an LLM-based code generator $G$, a prompt
$P\langle T,F\rangle$,
a reference code 
$R(P\langle T, F\rangle) = C^R_F$ for prompt $P$, and a set of formulas ${\mathcal{F}}_F$ containing syntactic variations of $F$ 
where for each $ F' \in {\mathcal{F}}_F$,
$ [\![ F' ]\!]  \equiv  [\![ F ]\!] $, 
let $ {\mathcal{F}}^{eq}_F \subseteq {\mathcal{F}}_F$ denote the set of
formulas such that for each $F' \in {\mathcal{F}}^{eq}_F$,
$[\![  G(P\langle T,F'\rangle) ]\!]
\equiv [\![  C^R_F ]\!]$. Then, 
the {\bf syntactic robustness degree} of $G$ with respect to $P$ and $\mathcal{F_F}$ is defined as:
\[
|{\mathcal{F}}^{eq}_F| / |{{\mathcal{F}}_F}| 
\]
where $|{\mathcal{F}}^{eq}_F|$ denotes the number of formulas in ${\mathcal{F}}^{eq}_F$ and $|{\mathcal{F}}_F|$ denotes the number of formulas in ${\mathcal{F}}_F$.
\label{robustness-degree}
\end{definition}
We report the syntactic robustness degree
as a percentage where 
100\% corresponds to the case
where ${\mathcal{F}}^{eq}_F ={\mathcal{F}}_F$. Note that the definition of syntactic robustness given in Definition~\ref{robustness} corresponds to the syntactic robustness degree of 100\% 
for all prompts and their corresponding semantically equivalent syntactic variations.








\section{Syntactic Mutations of Formulas} 
\label{sec:transformation_1}


In this paper, we consider prompts that include mathematical equations and expressions which can be univariate or multivariate.

\begin{figure}

\begin{equation*}
{
\begin{aligned} 
F \to \ & E = E \mid E\\ 
E \to \ & N \mid S \mid V \mid U \mid E + E \mid E - E 
 \mid E \times E \mid E / E \mid E\string^E \mid (E)\\ 
U \to \ & sin(E) \mid cos(E) \mid tan(E) \mid log(E) \mid ln(E) \mid V(E) \mid V ' (V) \\
N \to \ & 
 - N
 \mid [0-9]^+ 
 \mid [0-9]^+ . \ [0-9]^+
\\
S \to \ & a \mid b \mid c \mid d \mid e\\
V \to \ & x \mid y \mid z
\end{aligned}
}
\end{equation*}
\caption{Context-free grammar for mathematical formulas.}
\Description{Context-free grammar for mathematical formulas}
\vspace{-3mm}
\label{grammar}

\end{figure}


\paragraph*{Grammar for formulas.} The context-free grammar shown in Figure~\ref{grammar} captures the mathematical formulas we use, where $F$ denotes the start symbol which can be extended to either an equation ($E = E$) or an expression ($E$), $U$ denotes unary functions, $N$ denotes number literals, $S$ denotes coefficients, and $V$ represents the variables or functions. Note that, $V_1'(V_2)$ denotes the first derivative of function $V_1$ with respect to $V_2$.

Now, we discuss how we generate the set of formulas ${\mathcal{F}}_F$ that are syntactic variations of a given formula $F$. 


\begin{definition}
\small
A {\bf syntactic transformation} $ST$ is a function that maps a formula to another formula that is semantically equivalent, i.e., $ST:  {\mathcal{F}} \rightarrow {\mathcal{F}}$ 
such that for any formula $F \in {\mathcal{F}}$,
$[\![ F ]\!]  \equiv  [\![ ST(F) ]\!]$,  where 
${\mathcal{F}}$ denotes the set of formulas.
\label{transformation}
\end{definition}


We define 18 mutation types for equations and 12 mutation types for expressions that modify syntax while preserving the semantics of the mathematical formula. These mutations include commutative addition and multiplication, adding zero, multiplication by one, and variable renaming. Table~\ref{tab:muteq} presents a representative subset of the mutation rules for both expressions and equations, while the complete set is provided in the artifact~\cite{ourAnonRepo}.




Given a formula $F$, we generate the set of
syntactic mutations of $F$, called ${\mathcal{F}}_F$ (as described in Definition~\ref{robustness-degree}) by repeatedly applying mutations ($M(F), M(M(F))$, $M(M(M(F)))$,...)  to $F$, i.e.,
each $F' \in {\mathcal{F}}_F$ and by definitions of the mutations we introduced,
$ [\![ F' ]\!]  \equiv  [\![ F ]\!] $.


\begin{definition}
\small
Given a formula $F$ and its syntactic mutant $F'$, the {\bf syntactic distance} of $F$ and $F'$ is $n$ when $F' = M^n(F)$.
\label{distance}
\end{definition}
\noindent
I.e., the syntactic distance between $F$ and its mutant $F'$ is the number of mutations needed to mutate $F$ to $F'$.

\begin{table}[t]
\centering
\footnotesize
\caption{A representative subset of mutation rules for equations and expressions (where $I \to N \mid S$ and $0 \notin N$).}.
\label{tab:muteq}
\setlength{\tabcolsep}{12pt}
\renewcommand{\arraystretch}{1.7}
\begin{tabular}{|ll||ll|}
\hline
\multicolumn{2}{|c||}{Equation} & \multicolumn{2}{c|}{Expression}                            \\ \hline \hline
\multicolumn{1}{|l|}{$M_1$}    &  $\frac{E_1 = E_2}{E_1/I = E_2/I}$  & \multicolumn{1}{l|}{\multirow{2}{*}{$M_{1,2}$}} & \multirow{2}{*}{$\frac{E}{E \times I / I}$} \\ \cline{1-2}
\multicolumn{1}{|l|}{$M_2$}    &  $\frac{E_1 = E_2}{E_1\times I = E_2\times I}$  & \multicolumn{1}{l|}{}                  &                   \\ \hline
\multicolumn{1}{|l|}{$M_3$}    &  $\frac{E_1 = E_2}{E_1+I = E_2+I}$  & \multicolumn{1}{l|}{\multirow{2}{*}{$M_{3,4}$}} & \multirow{2}{*}{$\frac{E}{E + I - I}$} \\ \cline{1-2}

\multicolumn{1}{|l|}{$M_4$}    &  $\frac{E_1 = E_2}{E_1-I = E_2-I}$  & \multicolumn{1}{l|}{}                  &                   \\ \hline
\end{tabular}
\end{table}

\section{Attacks Targeting Syntactic Robustness}
\label{sec:attack}

We consider a black-box adversary aiming to reduce the syntactic robustness of an LLM in code generation tasks. The adversary can only apply semantics-preserving syntactic transformations to the input prompts and has no access to the model architecture or parameters. The goal is to induce incorrect or non-compiling code without changing the intended semantics. We assume the model is pre-trained and do not consider attacks on training data or deployment-stage injection. This threat model enables systematic analysis of LLM failure modes via adversarial robustness testing~\cite{zhou2024mathattack,randomperturbationattack25} and is applicable to automated code generation pipelines, where inputs are fed directly to LLM-based code generators~\cite{zhangetal2024codeagent} with minimal human oversight.



We use Algorithm~\ref{alg:profiling} to profile the impact of different mutations on syntactic robustness degree of LLMs. 
This algorithm assigns a score ($\mathcal{W}[M]$) for each mutation ($M$) based on the decrease in syntactic robustness from the original prompt ($P\langle T, F\rangle$) to a modified prompt ($P\langle T, M(F)\rangle$). Based on this profiling information, we also compute a threshold ($t_k$) which we use to choose a subset of the mutations that have the most impact on robustness ($\mathcal{M}_{Top-K}$).

\begin{algorithm}[t]
{\small
\caption{\textsc{Profiling($\mathcal{M}$, $\mathcal{P}$)} \\ 
\Comment{Decides effectiveness of each mutation for attack.} \\
\Comment{Calls function \textsc{Threshold} to choose a threshold to get top K mutations.}
}\label{alg:profiling}
\begin{flushleft}
  \textbf{Input:} $\mathcal{M}$: mutation set and $\mathcal{P}$: set of prompts to mutate.\\
  \textbf{Output:} $\mathcal{W}$: set of weights describing effect of each mutation type on syntactic robustness, $\mathcal{M}_{Top-K}$: A set of Top-K mutations.
\end{flushleft}
\begin{algorithmic}[1]
\For{$M \in  \mathcal{M}$}
    \State $\textit{robust\_score} \leftarrow \Sigma_{P \in \mathcal{P}} (|{\mathcal{F}}^{eq}_F| / |{{\mathcal{F}}_F}| - |{\mathcal{F}}^{eq}_{M(F)}| / |{{\mathcal{F}}_{M(F)}}|)$
    \State $\mathcal{W}[M] \leftarrow \textit{robust\_score}/|\mathcal{M}|$
\EndFor
\State $t_k \leftarrow $\textsc{Threshold}($\mathcal{W},k$)
\For{$M \in  \mathcal{M}$}
    \If{$\mathcal{W}[M] > t_k$}
    \State $\mathcal{M}_{Top-K} \gets \mathcal{M}_{Top-K} \cup \{ M \}$
\EndIf
\EndFor

\State \Return $\mathcal{W}, \mathcal{M}_{Top-K}$
\end{algorithmic}
}
\end{algorithm}

\begin{algorithm}[t]
{\small
\caption{\textsc{Attack($F$, $D$, $\mathcal{M}$)} \\ 
\Comment{Attacks by mutating a formula $F$ to distance $D$.}\\ 
\Comment{Calls function $\textsc{Random}$ which chooses mutations randomly based on the given weights.}}\label{alg:attacking}
\begin{flushleft}
  \textbf{Input:} $F$: formula, $D$: Distance (number of mutations to apply), $\mathcal{M}$: set of mutations\\
  \textbf{Output:} $F'$: mutated formula.
\end{flushleft}
\begin{algorithmic}[1]
\State $F' \leftarrow F$
\For{$D$}
    \If{Uniform}
     \State $M \leftarrow \textsc{Random}(\mathcal{M}, \left[\frac{1}{|\mathcal{M}|} \dots \frac{1}{|\mathcal{M}|} \right])$ 
     
    \ElsIf{Weighted}
        \State $M \leftarrow \textsc{Random}(\mathcal{M},\mathcal{W})$
    \ElsIf{Top-K}
          \State $M \leftarrow \textsc{Random}(\mathcal{M}_{Top-K},\left[\frac{1}{k} \dots \frac{1}{k} \right])$ 
         
    \EndIf
    \State $F' \leftarrow M(F')$ 
\EndFor
\State \Return $F'$
\end{algorithmic}
}
\end{algorithm}

Algorithm~\ref{alg:attacking} outlines our three attack strategies. Each strategy takes as input a formula
$F$ from a prompt and a syntactic distance $D$, and produces a mutated formula $F'=M^D(F)$. For these strategies, we take care to ensure that a formula at distance $D$ cannot be created by a series of mutations less than $D$. We construct the mutation space using a dynamic programming approach: We iteratively generate all unique formulas at distance 1, then distance 2 (excluding those already generated at smaller distances), and so on up to distance $D$.  We additionally introduce another attack strategy \textbf{NaiveUniform} which does not include this distance assurance---a mutation of distance $D$ with the \textbf{NaiveUniform} attack could potentially be created with a lower distance. 

\textbf{NaiveUniform} chooses randomly from mutation set, but can choose mutations that lead to a lower distance mutated formula.
The \textbf{Uniform} strategy in Algorithm~\ref{alg:attacking} selects a mutation uniformly at random from the set of all mutations. The \textbf{Weighted} strategy chooses mutations probabilistically with the weights calculated in \textsc{Profiling} based on each mutation's impact. Lastly, \textbf{Top-K} one selects a mutation randomly from top-k most impactful mutation candidates, $\mathcal{M}_{Top-K}$, which are identified by \textsc{Profiling}. For these three attack strategies in Algorithm~\ref{alg:attacking}, when we declare a formula $F'$ to have distance $D$ from $F$, we assure that $D$ is the shortest distance of mutations that can produce $F'$.

\begin{table}[h]
\centering
\footnotesize
\caption{A representative subset of reduction rules for equations (where $I \to N \mid S$ and $0 \notin N$)}.
\label{tab:reduceEquation}
\setlength{\tabcolsep}{12pt}
\renewcommand{\arraystretch}{1.7}
\begin{tabular}{|c|c|}
\hline
\multirow{2}{*}{$R_1$} & $\frac{E_1 = E_2}{E_1 - E_2 = 0}$ \\
 & Shift to L.H.S. \\ \hline
\multirow{2}{*}{$R_2$}  & $\frac{E + I - I}{E}\ \text{ ; } \ \frac{E_1 + I + E_2 - I = 0}{E_1 + E_2 = 0}\ \text{ ; } \ \frac{E_1 + I - E_2 - I = 0}{E_1 - E_2 = 0}$ \\
 & Remove redundant addition \\ \hline
\multirow{2}{*}{$R_3$}    & $ \frac{E - I + I}{E}\ \text{ ; } \ \frac{E_1 - I + E_2 + I = 0}{E_1 + E_2 = 0}\ \text{ ; } \ \frac{E_1 - I - E_2 + I = 0}{E_1 - E_2 = 0}$ \\
 & Remove redundant subtraction \\ \hline
\multirow{2}{*}{$R_4$}   & $ \frac{E \times I = 0}{E = 0}\ \text{ ; } \ \frac{E_1 \times I + E_2 \times I = 0}{E_1 + E_2 = 0}\ \text{ ; } \ \frac{E_1 \times I - E_2 \times I = 0}{E_1 - E_2 = 0}$ \\
 & Remove redundant multiplication \\ \hline
\multirow{2}{*}{$R_5$}   & $ \frac{E / I = 0}{E = 0}\ \text{ ; } \ \frac{E_1 / I + E_2 / I = 0}{E_1 + E_2 = 0}\ \text{ ; } \ \frac{E_1 / I - E_2 / I = 0}{E_1 - E_2 = 0}$ \\
 & Remove redundant division \\ \hline
\end{tabular}
\end{table}

\section{Prompt Pre-processing with Formula Reduction}
\label{sec:prepocessing}

To improve syntactic robustness, we add a preprocessing step where, instead of feeding a prompt with a mathematical formula to the LLM-based code generator as-is, we generate a reduced version of that formula. Then, we feed the reduced prompt with the same English text and the reduced mathematical formula to the LLM-based code generator to generate the target code. 

We focus on syntactic transformations that reduce the size of a formula, based on the assumption that shorter formulas are simpler and more likely to be handled correctly by LLM-based code generators. Accordingly, prompts with reduced formulas are expected to improve robustness against adversarial samples. We refer to these semantics-preserving syntactic transformations as {\em reductions}.

\begin{table}[h]
\centering
\footnotesize
\caption{A representative subset of reduction rules for expressions, given $I \to N \mid S$, where $0 \notin N$.}
\label{tab:reduceExpression}
\setlength{\tabcolsep}{12pt}
\renewcommand{\arraystretch}{1.7}
\begin{tabular}{|c|c|}
\hline
\multirow{2}{*}{$R_1$} & $\frac{E + I - I}{E}\ \text{ ; } \ \frac{E_1 + I + E_2 - I}{E_1 + E_2}\ \text{ ; } \ \frac{E_1 + I - E_2 - I}{E_1 - E_2}$ \\
 & Remove redundant addition followed by subtraction \\ \hline
\multirow{2}{*}{$R_2$}  & $\frac{E - I + I}{E}\ \text{ ; } \ \frac{E_1 - I + E_2 + I}{E_1 + E_2}\ \text{ ; } \ \frac{E_1 - I - E_2 + I}{E_1 - E_2}$ \\
 & Remove redundant subtraction followed by addition \\ \hline
\multirow{2}{*}{$R_3$}    & $\frac{E \times I / I}{E}$ \\
 & Remove redundant multiplication/division \\ 
 \hline
\end{tabular}
\end{table}

We define the size of a formula $|F|$ to be the number of terminal symbols in the formula as described in grammar shown in Figure~\ref{grammar}.

\paragraph*{Reductions} We formally define reductions as a type of syntactic transformations as follows:
\begin{definition}
\small
A {\bf reduction} $R$ is a syntactic transformation where for each 
formula $F \in {\mathcal{F}}$,
$|R(F)| \leq |F|$.
\label{mutation}
\end{definition}
\noindent
i.e., a reduction modifies the syntax of the formula without changing its semantics, while the size of the modified formula is either the same or less than the size of the original formula. 


We define 9 types of reductions for equations and a representative subset of these reduction rules are shown in Table~\ref{tab:reduceEquation} (rest are presented in the artifact~\cite{ourAnonRepo}).
Reduction rule R\textsubscript{1} positions all nonzero elements of the equation on one side of the equality operator. 
Reductions R\textsubscript{2} and R\textsubscript{3} show a series of possible applications for removing redundant additions and subtractions---specifically, the two final rules show removal of these redundant operations when a different additive or subtractive term intercedes the redundant operations. Reductions R\textsubscript{4} and R\textsubscript{5} show a series of possible applications for removing redundant divisions and multiplications---this is to show that this removal may not necessarily be from one singular $E$ comprising the full side of the formula but may also be from each individual additive or subtractive term on that side. For expressions, we introduce 7 reduction rules; we detail 3 in Table~\ref{tab:reduceExpression} and present the rest in the artifact~\cite{ourAnonRepo}.
Reductions R\textsubscript{1} and R\textsubscript{2} are similar to equation reductions R\textsubscript{2} and R\textsubscript{3}, to remove redundant additions and subtractions. Reduction R\textsubscript{3} removes redundant multiplications and divisions. The other reduction rules include canceling the addition of extra 0's, multiplication by 1's, and other similar simple reductions. (the full set of rules is in the artifact~\cite{ourAnonRepo}).
Besides equation reduction $R_1$ (which is applied once at the beginning), all reductions are applied in a loop until the formula cannot be simplified further. 

\section{Implementation and Experimental Design}
\label{sec:implementation}
\begin{figure*}[t]
\centering
\scalebox{0.9}{
\begin{tikzpicture}[
rectnode/.style={rectangle, minimum height=0.5, draw=black, thick, minimum width=0.1*\textwidth, text width=3cm, align = center},
parallelonode/.style={trapezium, trapezium left angle=60, trapezium right angle=120, draw=black, thick, minimum height=0.5, minimum width=0.1*\textwidth, text width=1cm, align = center},
diamondnode/.style={diamond, draw=black, thick, minimum height=0.5, minimum width=0.1*\textwidth, text width=3cm, align = center},
textnode/.style={trapezium, trapezium left angle=60, trapezium right angle=120, thick, minimum height=0.5, minimum width=0.1*\textwidth, text width=1cm, align = center},
]
\useasboundingbox (0,0) rectangle (0.9\textwidth,3);
\begin{scope}[use bounding box relative coordinates]
\node[parallelonode](prompt) at (0.02,0.5)[text width=0.8cm] {\footnotesize Prompt};
\node[rectnode](mutator) at (0.16,0.5)[text width=1.4cm, minimum height=30] {\footnotesize Formula Mutation};

\node[parallelonode](mutatedprompt) at (0.33,0.5)[text width=1cm] {\footnotesize Mutated Prompt};

\node[rectnode](llmgen) at (0.505,0.5)[text width=1.5cm, minimum height=25] {\footnotesize LLM-based Code Generator};

\node[parallelonode](gencode) at (0.65,0.25)[text width=0.8cm] {\footnotesize Generated Code};

\node[parallelonode](refcode) at (0.665,0.65)[text width=0.8cm] {\footnotesize Reference Code};

\node[diamondnode](equiv) at (0.825,0.5)[text width=1.2cm, minimum height=10] {\footnotesize Equivalent?};

\node[textnode](decision1) at (1,0.5)[text width=2cm] {\footnotesize Syntactically Robust};

\node[textnode](decision2) at (0.825,0)[text width=4cm] {\footnotesize Not Syntactically Robust};


\draw[->, thick] (prompt.east) -- (mutator.west);
\draw[->, thick] (mutator.east) -- (mutatedprompt.west);
\draw[->, thick] (mutatedprompt.east) -- (llmgen.west);
\draw[->, thick] (llmgen.south) -- ++(0.0,0.012) |- (gencode.west);
\draw[->, thick] (gencode.east) -| ++(0.035,0) |- (equiv.west);
\draw[->, thick] (equiv.east) -- node [below] {\footnotesize Yes} (0.93, 0.5);
\draw[->, thick] (equiv.south) -- node [left] {\footnotesize No} (0.825, 0.1);
\draw[->, thick] (refcode.north) -- ++(0.0,0.15) -- ++(0.16,0.01) -- (equiv.north);

 
\end{scope}
\end{tikzpicture}
}
\vspace{2mm}
\caption{Assessment of Syntactic robustness degree of an LLM (this workflow is applied in a loop for multiple mutations).}
\Description{A workflow of the syntactic robustness assessment technique of an LLM}
\label{fig:syntacticRobustness}
\end{figure*}

Figure~\ref{fig:syntacticRobustness} illustrates the workflow for assessing LLM-based code generators. We next describe our implementation and design.
\begin{table}[h]
\centering
\footnotesize
\caption{Summary of the code generation prompts}
\label{tab:promptSummary}
\resizebox{\columnwidth}{!}{
\begin{tabular}{|l|r||rr||rr|}
\hline
\multirow{2}{*}{\textbf{Prompt Types}} & \multicolumn{1}{l||}{\multirow{2}{*}{\textbf{Total}}} & \multicolumn{2}{l||}{\textbf{Classification 1}}                                    & \multicolumn{2}{l|}{\textbf{Classification 2}}                                      \\ \cline{3-6} 
                                       & \multicolumn{1}{l||}{}                                & \multicolumn{1}{l|}{\textbf{Expression}} & \multicolumn{1}{l||}{\textbf{Equation}} & \multicolumn{1}{l|}{\textbf{Reasoning}} & \multicolumn{1}{l|}{\textbf{Translation}} \\ \hline \hline
Solve                                  & 7                                                    & \multicolumn{1}{r|}{0}                   & 7                                      & \multicolumn{1}{r|}{7}                  & 0                                         \\ \hline
Evaluate                               & 7                                                    & \multicolumn{1}{r|}{7}                   & 0                                      & \multicolumn{1}{r|}{0}                  & 7   
                       \\ \hline
Differential                           & 6                                                    & \multicolumn{1}{r|}{1}                   & 5                                      & \multicolumn{1}{r|}{6}                  & 0     \\ \hline
MaTT~\cite{MaTTBenchmarkPaper}                                  & 6                                                    & \multicolumn{1}{r|}{5}                   & 1                                      & \multicolumn{1}{r|}{2}                  & 4                                         \\ \hline
MATH~\cite{MathBenchmarkPaper}                                 & 7                                                    & \multicolumn{1}{r|}{2}                   & 5                                      & \multicolumn{1}{r|}{6}                  & 1  
\\ \hline
Clover~\cite{clover}                                   & 19                                                    & \multicolumn{1}{r|}{0}                   & 19                                      & \multicolumn{1}{r|}{15}                  & 4 
\\ \hline \hline
\textbf{Total}                         & 52                                                   & \multicolumn{1}{r|}{15}                  & 37                                     & \multicolumn{1}{r|}{36}                 & 16                                        \\ \hline
\end{tabular}
}
\vspace*{-3mm}
\end{table}
\textbf{Prompts:} The prompts used in this paper are inspired from diverse scientific computing applications, including solving differential equations for molecular dynamics, as well as linear, polynomial, trigonometric, and logarithmic equations fundamental to techniques such as the Fourier Transform and Fast Fourier Transform (FFT)~\cite{Heckbert1998FourierTA}, shape function interpolation~\cite{zieli1992introduction}, and exponential growth or decay modeling. We also include prompts from established benchmarks~\cite{MathBenchmarkPaper,MaTTBenchmarkPaper} and contract-based programming~\cite{clover}. Table~\ref{tab:promptSummary} summarizes all prompt categories.
A \textit{solve} prompt includes an equation and asks to generate code that takes the equation’s coefficients as input and outputs the variable value that satisfies it. The \textit{evaluate} prompts instruct the LLM to generate a program that evaluates a mathematical expression given the values of all variables and coefficients. The \textit{differential} prompts involve manipulating differential formulas. The MATH~\cite{MathBenchmarkPaper} and Mathematical Topics Tree (MaTT)~\cite{MaTTBenchmarkPaper} datasets are designed to evaluate automated mathematical reasoning systems: MATH consists of challenging mathematics competition problems, while MaTT spans key topics across both pure and applied mathematics. The Clover~\cite{clover} dataset, in contrast, focuses on verifiable code generation using LLMs and contains prompts with formal specifications.

\textbf{Reasoning vs. Translation Prompts: } We categorize prompts into reasoning and translation types (Section~\ref{sec:overview}). Reasoning prompts require the LLM to engage in mathematical manipulation and reasoning to generate the appropriate code, as they involve complex tasks that need to be understood and processed based on the mathematical formula. In contrast, translation prompts allow the LLM to directly utilize the mathematical formulas that are in the prompts as part of the generated code, as these formulas can be directly translated into program expressions. Out of the 52 prompts, we have 36 reasoning and 16 translation prompts. 



\textbf{Programming language: } We chose C as the target language for our code generation prompts. C and C++ are extensively used for mathematical computing tasks in scientific computing due to their efficiency and performance in handling complex computations~\cite{numericalRecipeC,HighPerformaneC}. Recent studies have utilized these languages in LLM-based code generation for tasks such as differential equations, numerical methods, and advanced computer science problems~\cite{kashefi2023chatgpt,ChatGptForDEpresentation, GptForFiniteElementMethod,catir2025evaluating}.


\textbf{Mutated formula generation:} 
To measure the impact of syntactic transformations of a formula on the generated code, we generate up to 20 mutations per syntactic distance (1–5). At distance 1, the number is often fewer than 20 due to limited applicable mutation operators. Distance 0 corresponds to the original prompt, yielding a single instance. Across 52 prompts, this results in approximately 4,000 mutated prompts in total.

\textbf{Non-determinism of LLMs:} LLMs are inherently \linebreak non-deterministic. To address this, we experimented with various temperature, top\_p, and seed configurations (where applicable) and selected the most deterministic settings, detailed in our artifact~\cite{ourAnonRepo}. Additionally, we executed the LLM-based code generator five times per prompt to mitigate variability.


\textbf{Post-processing of the LLM responses:} Given a code generation prompt, the LLM responses typically contain the generated code and some English text explaining the generated code. Our implementation takes the generated response from an LLM, eliminates the non-code text, and converts it to a C file. The C files are then compiled into binaries using GCC~\cite{gcc}, which is automated in our pipeline. Any C code with syntactic errors is considered non-equivalent with respect to our reference code. 

\textbf{Differential Testing and Code Equivalence:} We implement reference solutions for each prompt and employ differential testing to evaluate the correctness and equivalence of the generated code. This involves generating 100 random inputs and comparing the outputs of the generated code with the reference solution. Numerical inputs are sampled from prompt-specific ranges. Our prompts also require LLM-generated outputs to follow a fixed format with either 6-digit or 2-digit precision, depending on the prompt type. For floating-point comparisons, outputs are considered correct if they fall within a predefined epsilon threshold relative to the expected result. This threshold mitigates rounding errors and relaxes our Definition~\ref{def:equivalence} of program equivalence. The epsilon values are provided in our artifact~\cite{ourAnonRepo}. When differential testing detects non-equivalence between generated code and reference solution, non-equivalence is guaranteed. However, differential testing cannot guarantee equivalence since it explores only a subset of the input space. In a manual review of 100 generated programs, we found no cases where non-equivalent programs were incorrectly classified as equivalent.

\textbf{Hyperparameter for Attack Strategies:} For the Top-K attack strategy, we have chosen the top 50\% mutations with the most impact on the syntactic robustness degree of LLM (calculated based on Gpt-3.5). The other strategies choose the mutations randomly based on the weights (Algorithm~\ref{alg:attacking}).
\textbf{Benchmark Models:} 
Our evaluation includes Gpt-3.5, Gpt-4o~\cite{gpt3.5turbo,openai2024gpt4}, Llama-3.1-70B~\cite{dubey2024llama}  and CodeLlama~\cite{roziere2023code}. Gpt-3.5 offers cost-efficiency, while Gpt-4 consistently performs well in both code generation and mathematical reasoning~\cite{NamingEffectPaper,ahn2024large}. Llama models are included for their open-source accessibility and low cost. We also evaluate o4-mini~\cite{o4-mini}, a reasoning model that balances performance with reduced computational overhead compared to other OpenAI reasoning models like o3, o1-pro. We experiment with Chain-of-thought (CoT)~\cite{wei2022chain} and Deeply Understanding Problems (DUP)~\cite{zhong2024achieving} approaches as they are zero-shot prompting techniques that are generalizable, and cost-efficient. For CoT, we use trigger-based prompts (e.g., “Let’s solve this step-by-step”) to preserve automation~\cite{kojima2022large}. DUP, which addresses semantic misunderstandings for mathematical reasoning, has achieved state-of-the-art results on arithmetic tasks~\cite{sotapaperswithcode} and is adapted here for mathematical code generation. In code generation accuracy, Gpt-3.5 and Gpt-4 frameworks outperform others, achieving $\geq$ 85\% on HumanEval~\cite{LLMTrainedonCode}, a trend reflected in our results; thus, framework-based techniques are applied only to Gpt models, excluding Llama models due to lower performance.


\textbf{Artifact:} Our artifact, including implementation, datasets, results, mutation and reduction rules, is available in GitHub~\cite{ourAnonRepo}.

\section{Experimental Evaluation}
\label{sec:experiments}


We address four research questions:
\begin{description}
\item \textbf{RQ1}: Assessment 1: Does LLM-based code generation achieve syntactic robustness?
\item \textbf{RQ2}: Assessment 2: Do LLMs show different levels of syntactic robustness for translation and reasoning prompts in code generation?

\item \textbf{RQ3}: Exploitation: Do the attacks help in decreasing the syntactic robustness of the LLM-based code generation?


\item \textbf{RQ4}: Mitigation: Does prompt pre-processing with formula reduction improve the syntactic robustness of code generation and help in mitigating our attacks?
\end{description}

Below we present the results of our experiments and discuss the four research questions. For our assessment (RQ1-2), we use the Gpt-3.5, Gpt-4o, Llama-3.1-70B, CodeLlama, Gpt-based framework models and we use the 33 prompts from Table~\ref{tab:promptSummary} (Row 1-5). For RQ3, we have chosen Gpt-3.5 as it performs better than open source models and cheaper than other Gpt-based models. To evaluate our reduction technique (RQ4), we have extended our dataset to Clover dataset and also included o4-mini model. 





\begin{figure}
    \centering    
    \small
    \includegraphics[width=.85\linewidth]{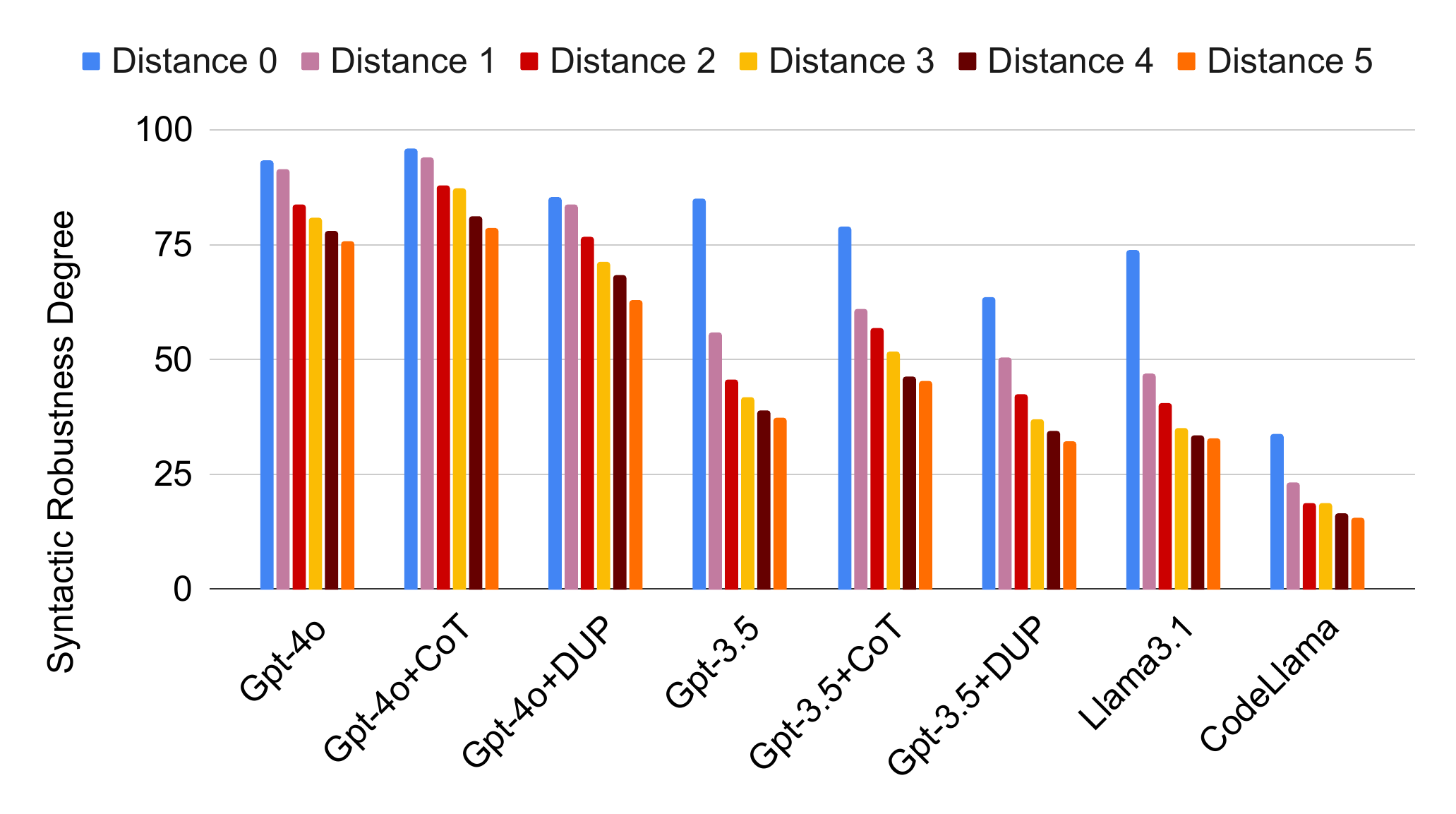}
     \caption{Syntactic Robustness Degree Vs. Distance for LLMs}
     \Description{A column chart showing the syntactic robustness degree vs. distance for LLMs}
    \label{fig:RQ1-2}
\end{figure}
\begin{figure}[tp]
    \centering
\includegraphics[width=0.85\linewidth]{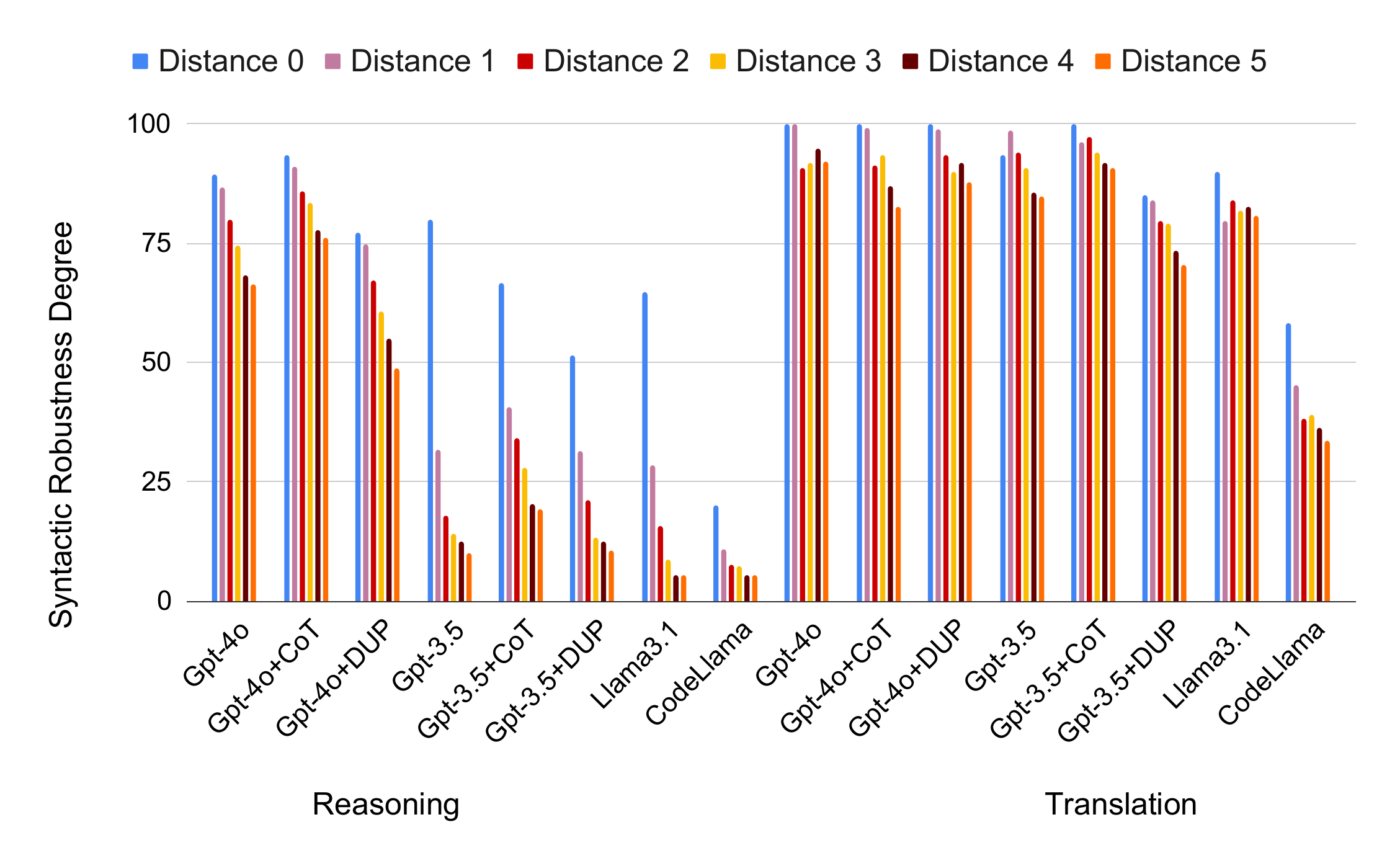}
    \caption{Syntactic robustness degree of reasoning prompts (left 8) and translation prompts (right 8)}
    \Description{A column chart showing the syntactic robustness degree of reasoning and translation prompts}
    \label{fig:RQ4-1}
\end{figure}
\textbf{RQ1: Assessment 1: }
\label{rq1} 
Figure~\ref{fig:RQ1-2} illustrates the syntactic robustness percentages for all foundational and framework models across different syntactic distances, based on the average robustness scores across the 33 prompts used. Distance 0 shows the robustness of code generation for the original formula, while Distance 1 through Distance 5 shows the average syntactic robustness degrees for the mutated variations at each respective syntactic distance. As shown in Figure~\ref{fig:RQ1-2}, the syntactic robustness degree decreases consistently across all models as syntactic distance increases. Despite varying performance levels, none of the models demonstrate complete syntactic robustness. Gpt-4o+CoT performs best, with an average syntactic robustness of 87.4\%, followed by Gpt-4o with 83.87\%. 

Moreover, Figure~\ref{fig:RQ1-2} shows average syntactic robustness percentages across mutation distances from 0 to 5 and clearly demonstrates that as the number of mutation increases, syntactic robustness decreases across all foundational and framework-based models. Notably, the rate of decline varies among different LLMs. 
Gpt-4o+CoT shows the slowest decline in syntactic robustness degree per mutation distance: -3.45\% on average. 
Gpt-3.5 experiences a more significant drop of -9.52\%, as its syntactic robustness degree with the original prompt is higher than the robustness degree of mutated prompts. 
In summary, the syntactic robustness decreases for all models as mutation distance increases.
Thus, we conclude that \textbf{LLMs do not achieve syntactic robustness} and \textbf{increasing number of mutations (equivalently, increasing syntactic distance) leads to a decrease in syntactic robustness}. 

\begin{figure}
    \centering
    \includegraphics[width=0.85\linewidth]{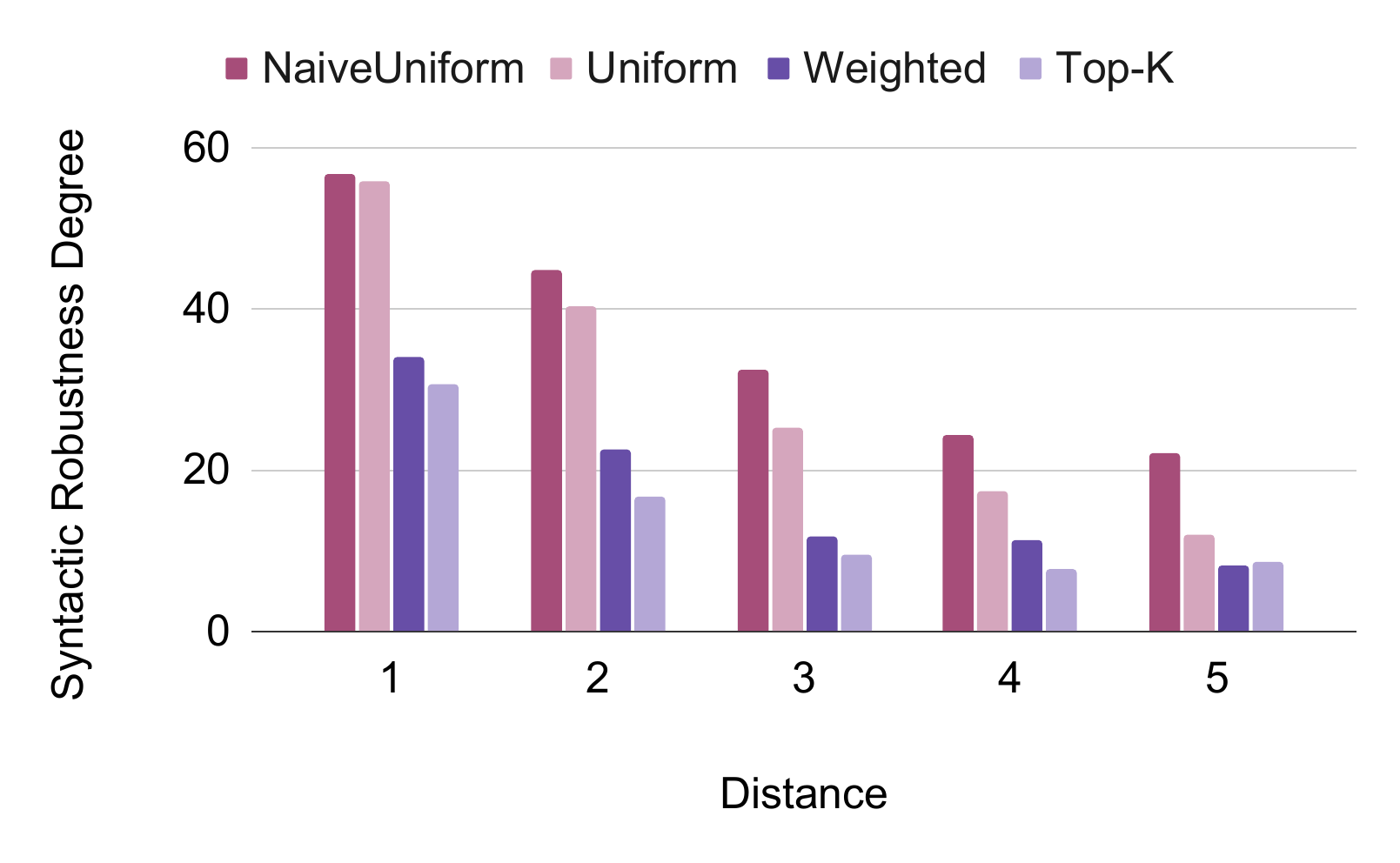}
    \caption{Syntactic robustness degree of different attacks for reasoning prompts (Gpt-3.5)}
    \Description{A column chart showing syntactic robustness degree of different attacks for reasoning prompts}
    \label{fig:reasoning_attack}
\end{figure}

\begin{figure}
    \centering
    \includegraphics[width=0.85\linewidth]{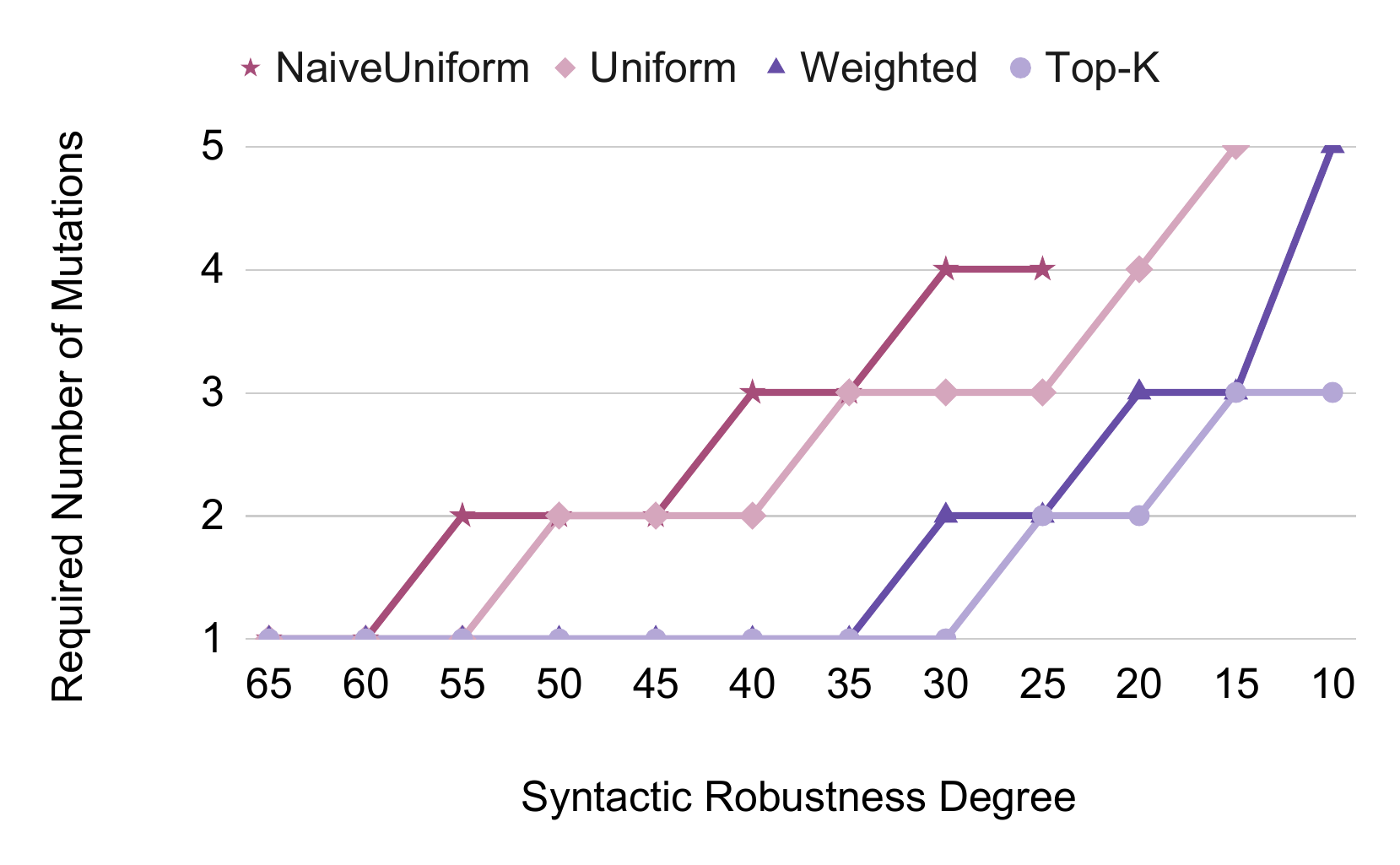}
    \caption{Required number of mutations for different attacks}
    \Description{}
    \label{fig:attacks_mutations}
\end{figure}

\textbf{RQ2: Assessment 2: } 
Figure~\ref{fig:RQ4-1} shows the average syntactic robustness of all reasoning prompts (left side) and translation prompts (right side) across eight models. The results indicate that translation prompts consistently exhibit higher robustness than reasoning prompts. As mutation distance increases, robustness declined more steeply for reasoning prompts. On average, reasoning prompts achieve a syntactic robustness of 42.86\%, compared to 83.80\% for translation prompts. The robustness drops by approximately 30\% for reasoning prompts and 8\% for translation ones relative to their original forms. 
Based on our experimental evaluation, we conclude that \textbf{LLMs exhibit different levels of syntactic robustness across translation and reasoning prompts, with greater robustness for translation prompts}.
\textbf{RQ3: Exploitation: }  As described in RQ2, translation prompts are more robust than the reasoning prompts. Attacks on the translation prompts also show similar results (details in artifact~\cite{ourAnonRepo}).
Thus, to answer this research question, we only focus on reasoning prompts. Figure~\ref{fig:reasoning_attack} shows the impact of our attack strategies on Gpt-3.5 model. It illustrates that for all attack strategies, increasing syntactic distance decreases the syntactic robustness degree, which follows the findings in RQ1. On average, across all distances, the syntactic robustness degree is 36.13\%, 30.17\%, 17.54\% and 14.64\% respectively for NaiveUniform, Uniform, Weighted and Top-K. The Weighted and Top-K attack strategies are most effective in reducing syntactic robustness degree across all tested distances. 
In Figure~\ref{fig:attacks_mutations}, we show how many mutations are required for different attack strategies to reduce the syntactic robustness degree to a target level (x-axis). 
With NaiveUniform, 4 mutations are required to reduce syntactic robustness to 25\% and more than 5 would be necessary for further reduction. On the other hand, similar syntactic robustness can be achieved with a lower number of mutations for the other attacks (3 for Uniform and 2 for Weighted and Top-K). To reduce the syntactic robustness down to 10\%, Weighted requires 5 mutations whereas Top-K requires only 3 mutations. Thus, Figure~\ref{fig:attacks_mutations} illustrates that Top-K is most effective in getting the highest reduction in syntactic robustness with the least number of mutations. 
Our experimental evaluation shows that \textbf{our attack strategies are effective in decreasing the syntactic robustness degree}.






\textbf{RQ4: Mitigation:}  
\begin{figure}[t]
    \centering
\includegraphics[width=.85\linewidth]{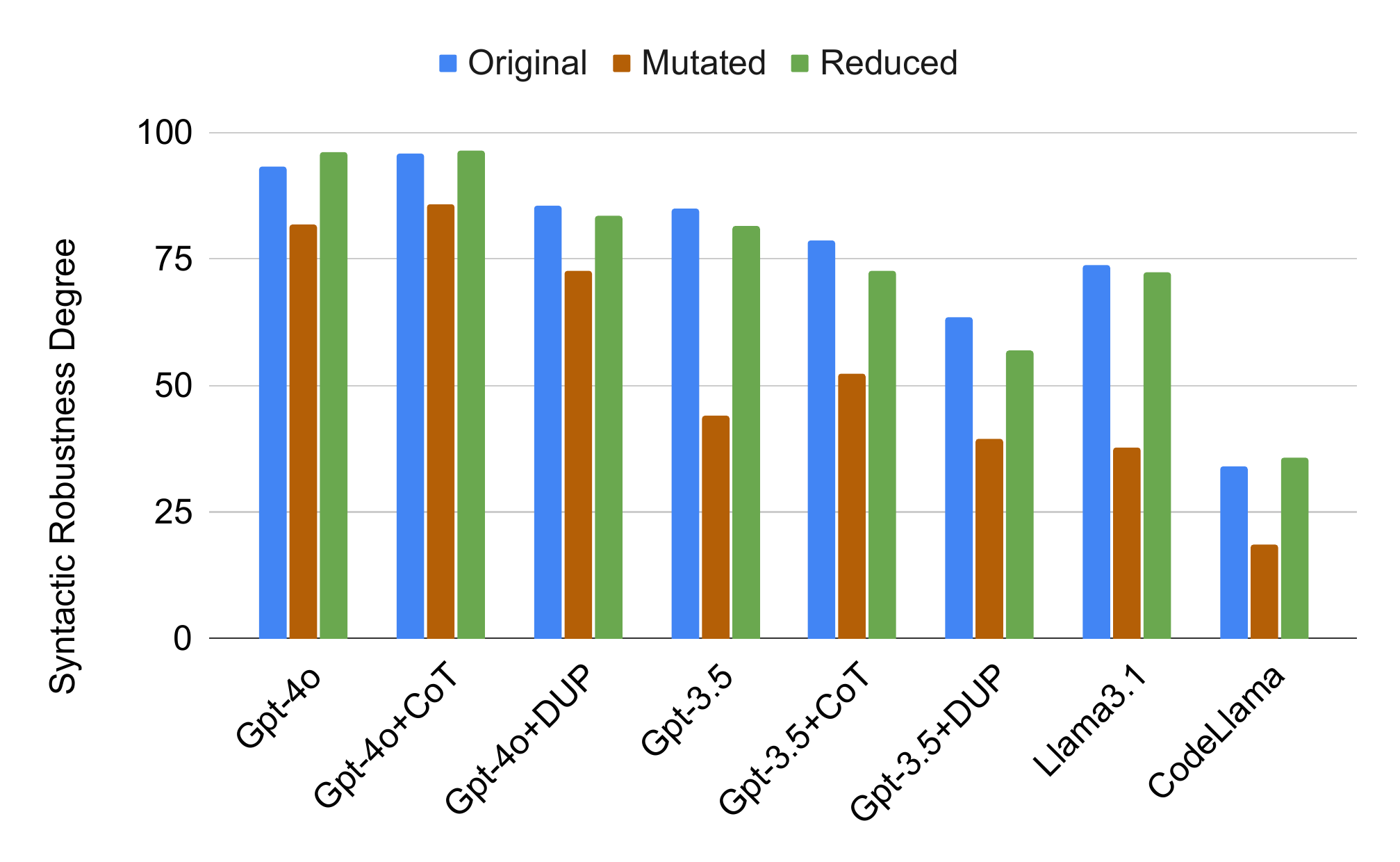}
    \caption{Syntactic robustness degree with reduced form}
    \Description{A column chart showing syntactic robustness degree of original, mutated and reduced form}
\vspace*{-4.8mm}
    \label{fig:RQ3-1}
\end{figure}
\begin{figure}
    \centering
    \includegraphics[width=0.85\linewidth]{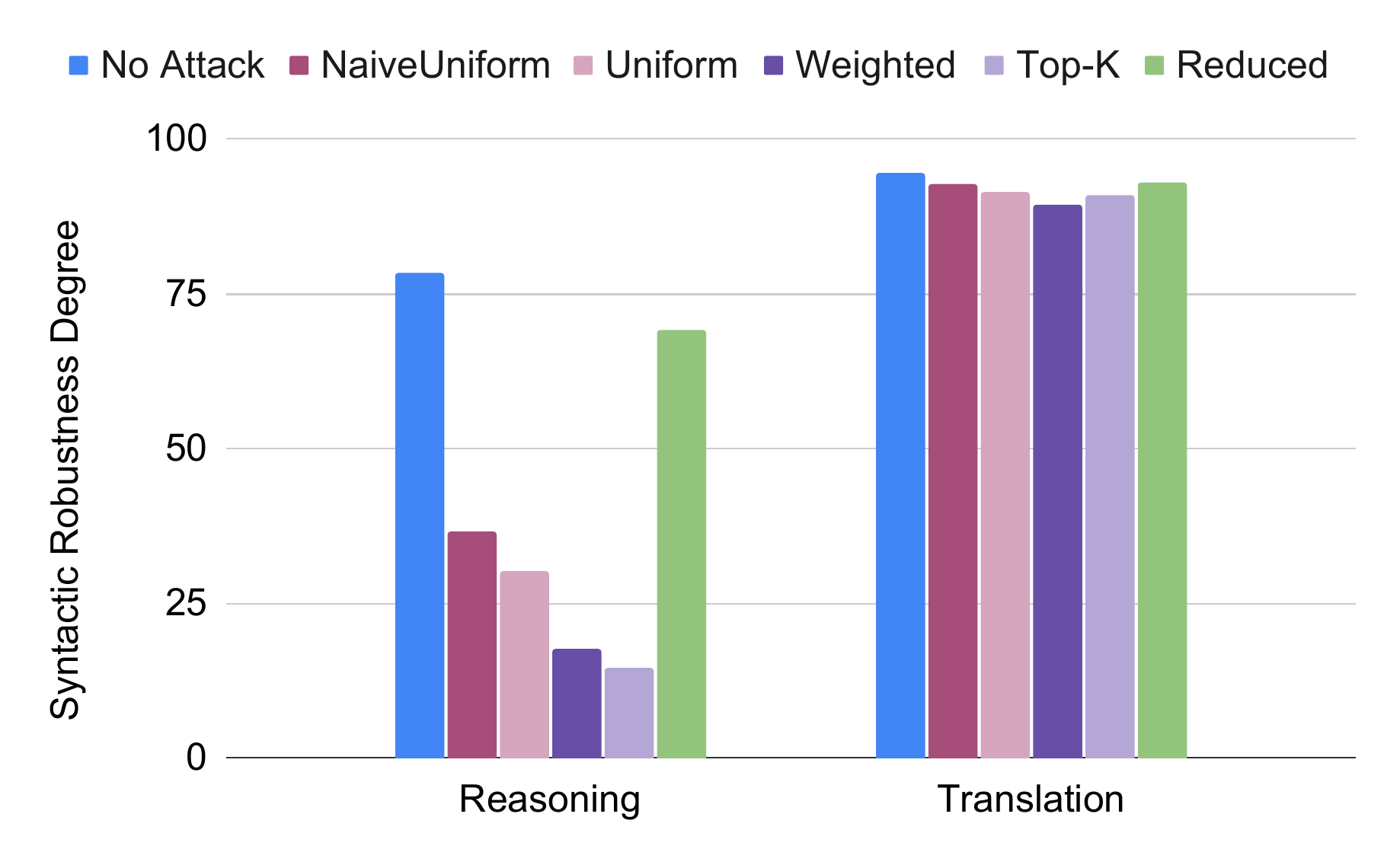}
    \caption{Mitigation of attack strategies (Gpt-3.5)}
    \Description{A column chart showing syntactic robustness degree of different attacks and reduced form}
    \label{fig:impact_attacks}
\end{figure}
\begin{figure} [t]
    \centering
    \includegraphics[width=0.8\linewidth]{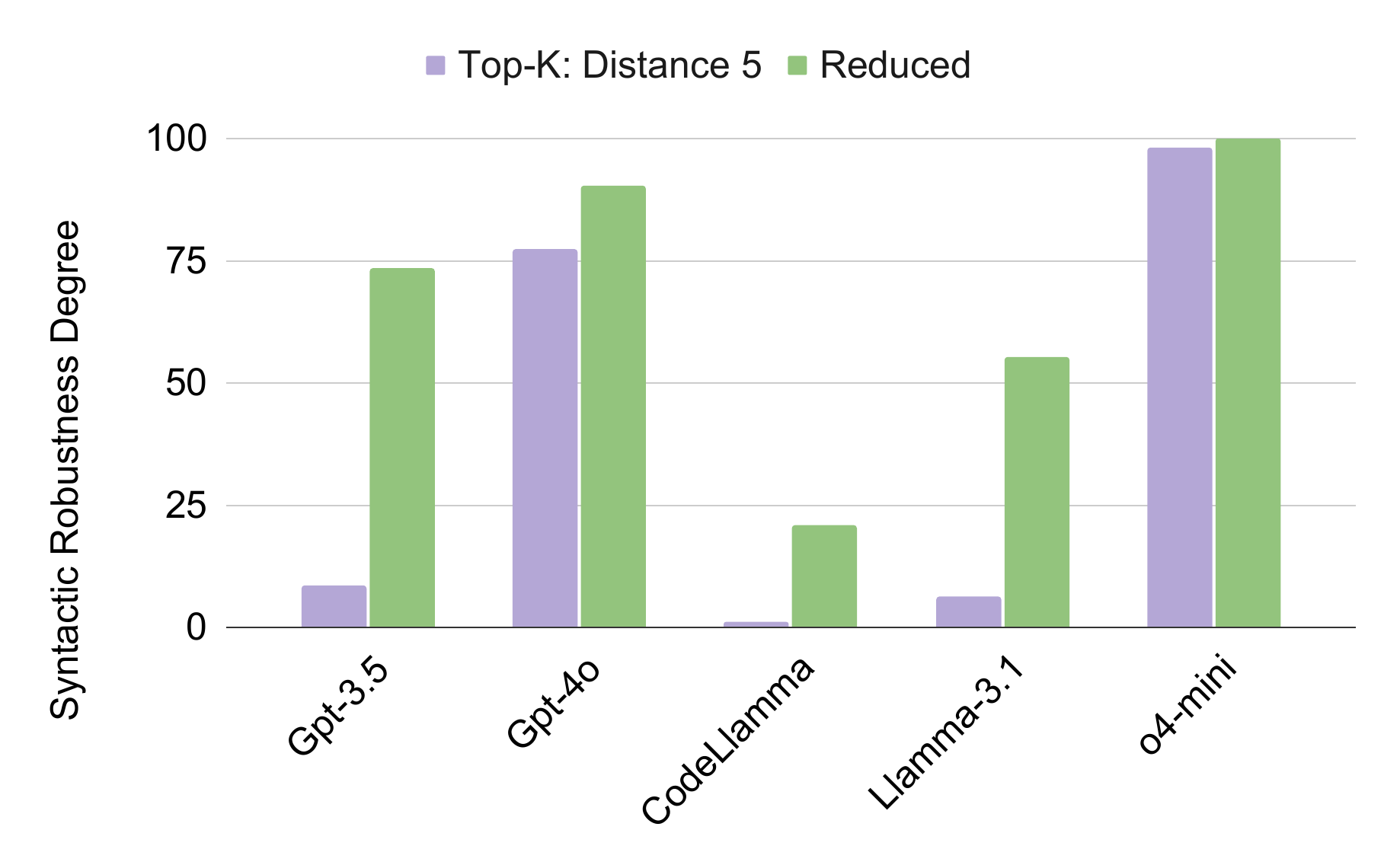}
    \caption{Impact of Top-K attacks on reasoning prompts at distance 5}
    \Description{A column chart showing the impact of Top-K attacks on reasoning prompts at distance 5 along with reduced form}
    \label{fig:llm_a4_d5}
\end{figure}
\begin{figure} [t]
    \centering
    \includegraphics[width=0.81\linewidth]{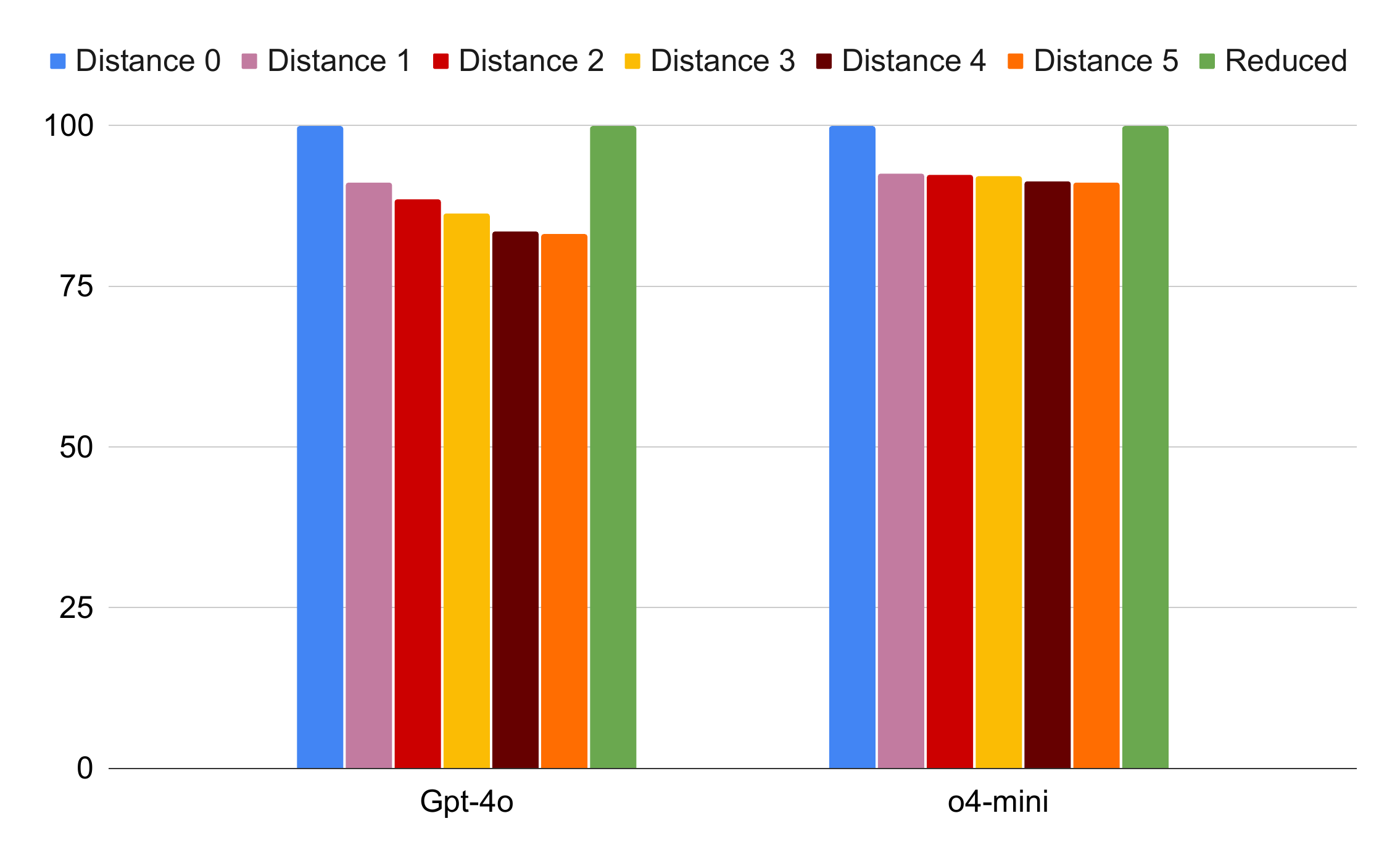}
    \caption{Top-K attack for Clover Dataset}
    \Description{A column chart showing Top-K attack on Clover dataset on Gpt-4o and o4-mini models}
    \label{fig:llm_a4_all_dist}
\end{figure}
We demonstrate the mitigating effect of applying our reduction rules to mutated formulas before querying the LLM-based code generator. Figure~\ref{fig:RQ3-1} presents the syntactic robustness of the models for the original formulas, the average robustness for mutated formulas, and the robustness of the reduced prompts. Across all foundational and framework models, the reduced prompts exhibit robustness degrees comparable to the original prompts and show notable improvement over the mutated prompts. The improvement in syntactic robustness when comparing reduced prompts to mutated prompts is substantial, with increases between 10.63\% and 37.55\%, resulting in an overall average improvement of 20.37\%. Among all foundational and framework models, the reduced prompts for Gpt-4o+CoT perform the best on average for all the prompts we have used.

For Gpt-4o and Gpt-4o+CoT, the reduced prompts show improvements of 2.73\% and 0.6\%, respectively, even over the original prompts. Moreover, CodeLlama achieves an improvement of 5.1\% for reduced prompts compared to the original ones. In contrast, other models show a slight degradation in performance with the reduced prompts compared to the original prompts. 

Figure~\ref{fig:impact_attacks} shows the syntactic robustness degrees of different attacks along with the original and our reduced prompts for both reasoning and translation prompts in Gpt-3.5. It is evident that our pre-processing step helps in mitigating the impact of attacks, especially for reasoning prompts. Moreover, in Figure~\ref{fig:llm_a4_d5}, we have shown the results for the \textsc{Top-K} attack at distance 5 and the robustness for the corresponding reduced prompts across different LLMs (including o4-mini). In all cases, reduction improves the syntactic robustness. Even though o4-mini is very robust against the attack, there is a 1.95\% drop in robustness which can be mitigated using our reduction step. Figure~\ref{fig:llm_a4_all_dist} also shows the result across different distances for Top-K attack on Gpt-4o and o4-mini on Clover dataset. This result on a different dataset also follows the trend of decreasing robustness with distances which was successfully mitigated by our reduction steps. 
Overall, our reduction rules are effective in pre-processing the prompts for reliable code generation, and the experimental results confirm that \textbf{applying our pre-processing step to simplify formulas enhances the syntactic robustness of LLM-based code generation}.

\paragraph{Discussion} All foundational models were evaluated using the same prompts from our dataset. However, the framework models, such as Gpt+CoT and Gpt+DUP, require additional text in the prompts. Although this extra text does not alter the instructions about the input and output format of the code, we observed that the framework models, particularly Gpt-3.5+DUP and Gpt-3.5+CoT, often fail to follow the given instructions. For instance, Gpt-3.5+DUP produced only the instruction to generate code in the comment section of the C program in 22 cases, while Gpt-3.5+CoT did the same in 3 cases. Additionally, in 18 cases, Gpt-3.5+DUP generated code without incorporating the input parameters, which disrupts the automated differential testing process. This issue is even more pronounced in Llama-based models, where code is generated without user input processing in approximately 150 cases for Llama-3.1-70B and 600 cases for CodeLlama. These factors negatively impact the overall performance of Llama and the framework-based models in code generation. We also observe that Llama-3.1 and CodeLlama often choose to present results in other languages such as C++ and Python despite clear instructions to use C, which lowers the syntactic robustness. Although we selected o4-mini for its efficiency over larger reasoning models like o3 and o1-pro, our experiments reveal that it still takes approximately $4\times$ more time than Gpt-4o and about $10\times$ more time than Gpt-3.5 for comparable tasks. 

\paragraph{Threats to Validity} We discuss possible internal and external threats to validity and the efforts we have made to mitigate them. 
{\em Internal:} There are a few possible internal threats to validity within our experimental design. First, there is the possibility that the Gpt API (for Gpt-3.5, Gpt-4o, and o4-mini) may be learning from prior requests we send. However, we believe this is not the case, as we have switched APIs during these experiments multiple times and found no observable difference between code returned from a previously-used API and code returned from a fresh one. Second, we note that differential testing cannot prove equivalence, and due to needing an epsilon value in 31 prompts to avoid incorrectly marking nonequivalence due to floating point error, it is possible some incorrect results could be marked correct. This is a possible risk for any testing-based equivalence checking approach. However, we have manually investigated a subset of LLM generated programs and found no case where differential testing falsely marked two non-equivalent codes as equivalent. Moreover, when we declare generated code non-equivalent, non-equivalence is guaranteed since we produce a test case that demonstrates the non-equivalence. 

\noindent {\em External:} Analyzing all formulas or mutations is not possible. However, we believe that our evaluation shows syntactic robustness issues of LLMs for an important class of formulas and mutations. Furthermore, our evaluation and attack approaches can be applied to different classes of formulas and syntactically equivalent forms of those formulas using the framework we present in this paper. 


\paragraph{Limitations} Our work focuses on the robustness of LLMs specifically for code generation prompts involving mathematical formulas and constraints. Consequently, our transformation and reduction techniques are also limited to prompts containing such mathematical formulas. However, mathematical computation–based code generation is fundamental to many domains, including scientific computing and contract-based programming. While our benchmark consists of a limited number of prompts, we believe they represent a diverse set of prompts with mathematical formulas and can be further extended. Additionally, while we evaluate code generation for the C programming language, our approach can be extended to assess other LLMs for different programming languages.

\section{Related Work}
\label{sec:relwork}

Prior work on LLM-based code generation primarily focuses on improving performance on software engineering tasks~\cite{LLMTrainedonCode,li2022competition,nijkamp2022codegen,xu2022systematic}, often evaluated on benchmarks such as HumanEval, MBPP, and APPS~\cite{LLMTrainedonCode,austin2021program,hendrycks2021measuring}. In parallel, substantial research studies mathematical problem solving with LLMs~\cite{he2023solving,uesato2022solving,zhong2024achieving}, using datasets such as GSM8K, SVAMP, and MultiArith~\cite{cobbe2021training,patel2021nlp,roy2016solving}. Recent work in scientific computing and numerical analysis~\cite{kashefi2023chatgpt,llm4ScientificDiscovery,shojaee2024llm,du2024llm4ed} shows that these capabilities can be combined to solve complex tasks, including contract programming~\cite{clover}. However, prior work does not examine the robustness of LLM-based code generators under syntactic variations, which is the focus of this work.

Robustness has been studied for classification and regression neural networks, using techniques such as symbolic reasoning~\cite{katz2019marabou,bunel2020branch,sun2018concolic}, abstraction~\cite{singh2019abstract,wang2018formal}, and testing/fuzzing~\cite{baluta2021scalable,xie2019deephunter}.

Prior work has examined the correctness of LLMs for code generation tasks~\cite{CopilotCodewhispererChatGPT,ahmed2023improving,EvalPlus,borji2023categorical,dinh2024large,spiess2024calibration,he2024beyond,wang2024examining}. Other studies investigate robustness against non-determinism in LLMs~\cite{nondeterminism}, perturbations in natural language descriptions~\cite{robustnessCopilot,jiang2022discovering}, and variations in coding components~\cite{yan2023coco,stackOverReplacementRobustness,buscemi2023comparative,xu2022systematic,rabbi2025multi}. ReCode~\cite{recode} introduces a robustness evaluation benchmark for code generation by perturbing prompts, including both natural language components and code segments via partial code syntax refactoring in code completion settings. Semantic preservation in ReCode relies on human annotation. In contrast, our work evaluates robustness through perturbations of mathematical formulas that are unambiguously semantics-preserving and do not require human annotation. Moreover, ReCode’s prompts, transformations, and datasets do not target mathematical reasoning and do not include prompts containing mathematical formulas. Robustness has also been studied for mathematical problem solving by LLMs~\cite{anantheswaran2024investigating,li2024gsm,kumar2021adversarial,zhang2024mario,zhou2024mathattack,hao2025investigation} which are not focused on code generation by LLMs. Compared to prior work and surveys~\cite{Yang2024RobustnessSP,fan2023large}, our work formally defines syntactic robustness and focuses on prompts containing mathematical formulas, targeting the robustness of LLM-based code generation.

Our syntactic transformation approach, which mutates mathematical formulas, is conceptually similar to metamorphic testing~\cite{chen2020metamorphic}, a property-based technique also applied to LLMs~\cite{applis2021assessing, applis2023searching}. For example, LAMPION~\cite{applis2021assessing} generates equivalent code snippets to test the robustness of LLM-based program analysis models. In contrast, we target LLM-based code generation.

Recent multi-agent code generation frameworks like AgentCoder, LDB, and MapCoder~\cite{huang2023agentcoder,zhong2024ldb,islam2024mapcoder} achieve over 90\% accuracy on benchmarks such as HumanEval and MBPP~\cite{sotapaperswithcode}, but require multiple LLM queries per prompt~\cite{kapoor2024ai, aileaderboards}. Studies show that when both accuracy and cost are considered, these frameworks do not outperform baseline foundational models~\cite{kapoor2024ai}, and they also rely on input-output test cases~\cite{hendrycks2021measuring, khan2023xcodeeval}. In contrast, our work focuses on prompt-engineering frameworks that rely solely on prompts.

Beyond multi-agent methods, significant progress has been made with prompt engineering frameworks~\cite{improvingPromt,marvin2023prompt,li2023towards,doderlein2022piloting,he2023controlling,white2023chatgpt,li2023skcoder,li2023enabling,ahmed2023improving,jiang2023selfevolve,zhang2023self,mu2024clarifygpt}. CoT has been applied to code generation by decomposing problems into sequential, branch, and loop structures~\cite{shao2023synthetic, li2023structured}. Other reasoning-oriented prompting strategies, such as DUP and AceCoder~\cite{zhong2024achieving,li2023acecoder,jiang2023self,anantheswaran2024investigating,didolkar2024metacognitive,wu2024mathchat}, also achieve strong performance, with DUP attaining state-of-the-art results on arithmetic reasoning benchmarks~\cite{sotapaperswithcode}. Since our dataset combines mathematical reasoning and code generation, we evaluate both CoT and DUP.

Prior robustness attacks on deep code models~\cite{qu2024survey} primarily target misclassification in tasks like authorship attribution~\cite{gao2023discrete, liu2021practical}. In contrast, we focus on LLM-based code generators, which differ from classifiers. While black-box prompt attacks~\cite{xu2023llm,zhou2024mathattack} exist, they do not address code generation. Other work has explored code completion or generation attacks using adversarial comments (INSEC~\cite{jenkoblack}) or non-functional code changes~\cite{randomperturbationattack25}. Our approach instead mutates mathematical formulas to produce semantically equivalent prompts, exposing failures in syntactic robustness.

There are also works which focus on training strategies or architectural changes for improving the robustness of LLM code generations~\cite{dingsemcoder,chakraborty2022natgen,pei2024exploiting,zhang2024codefort}. Our approach is distinct as we focus on black-box prompt pre-processing for improving this robustness, without altering or retraining the models.

\section{Conclusion}
\label{sec:conclusion}

The use of LLM-based code generation is expanding in domains involving mathematical computations. In this paper, we expose limitations of state-of-the-art LLMs when handling prompts with mathematical formulas under semantic-preserving syntactic mutations. We formalize this concept as syntactic robustness and evaluate multiple benchmark models using both original and mutated prompts. Our results show a significant decrease in robustness as syntactic distance increases, with reasoning-intensive prompts particularly affected. We further design and compare attack strategies that effectively reduce robustness. To mitigate this issue, we introduce a formula reduction-based prompt pre-processing technique that consistently improves robustness across all evaluated models.

\balance

\bibliographystyle{ACM-Reference-Format}


\end{document}